 \documentclass[conference,a4paper,10pt]{IEEEtran}
 \usepackage{amsfonts}
 \IEEEoverridecommandlockouts
\usepackage{epsfig}
\usepackage{graphicx}
\usepackage{psfig}
\usepackage{subcaption}
\usepackage{epsf}
\usepackage{epstopdf}
\usepackage{bbm}
\usepackage{colortbl}
\usepackage[most]{tcolorbox}
\usepackage{tikz}
\usetikzlibrary{arrows.meta,shapes,decorations,automata,backgrounds,petri,topaths,calc,mindmap,trees,positioning,chains,arrows}
\usepackage{relsize}
\usepackage{pgfplots}
\pgfplotsset{compat=1.14}
 \usepackage{algorithm2e}
\usepackage{algpseudocode}
\usepackage{amssymb}
\usepackage{float}
\usepackage{multirow}
\usepackage{pbox}

\usepackage{amsmath}
\allowdisplaybreaks[4]
\usepackage{booktabs}
\usepackage{fancyhdr}

\usepackage{epstopdf, cite,color}
\usepackage{amsthm}
\usepackage{amssymb}
\newcommand{\bc}{\begin{center}}
	\newcommand{\ec}{\end{center}}
\newcommand{\be}{\begin{equation}}
\newcommand{\ee}{\end{equation}}
\newcommand{\bea}{\begin{eqnarray}}
\newcommand{\eea}{\end{eqnarray}}

\begin{document}

\title{Rethinking On-Device LLM Reasoning: Why Analogical Mapping Outperforms Abstract Thinking for IoT DDoS Detection}
% From Incompetent to Expert: Enabling On-Device LLMs for IoT Attack Detection via Few-Shot Retrieval-Augmented Generation}

\author{William Pan\textsuperscript{\S}, Guiran Liu\textsuperscript{\S}, Binrong Zhu\textsuperscript{\S}, Qun Wang\textsuperscript{\S}, Yingzhou Lu \textsuperscript{{\dag}}, Beiyu Lin \textsuperscript{{\ddag}}, Rose Qingyang Hu\textsuperscript{{\pounds}} 

    \\

	 \textsuperscript{\S}Department of Computer Science, San Francisco State University, San Francisco, CA, 94132\\

\textsuperscript{{\dag}}School of Medicine, Stanford University, USA\\
  \textsuperscript{{\ddag}} Department of Computer Science, University of Texas Dallas, Dallas, USA\\
  \pounds Bradley Department of Electrical and Computer Engineering, \\Virginia Polytechnic Institute and State University,  Blacksburg, VA 24061\\
    Emails:  
       William910122@gmail.com, claudqunwang@ieee.org, gliu@sfsu.edu, bzhu2@sfsu.edu,\\lyz66@stanford.edu,
	beiyu.lin@utdallas.edu,
    rosehu@vt.edu
    }
\maketitle

% \IEEEpeerreviewmaketitle
\begin{abstract}

The rapid expansion of IoT deployments has intensified cybersecurity threats, notably Distributed Denial of Service (DDoS) attacks, characterized by increasingly sophisticated patterns. Leveraging Generative AI through On-Device Large Language Models (ODLLMs) provides a viable solution for real-time threat detection at the network edge, though limited computational resources present challenges for smaller ODLLMs. This paper introduces a novel detection framework that integrates Chain-of-Thought (CoT) reasoning with Retrieval-Augmented Generation (RAG), tailored specifically for IoT edge environments. We systematically evaluate compact ODLLMs, including LLaMA 3.2 (1B, 3B) and Gemma 3 (1B, 4B), using structured prompting and exemplar-driven reasoning strategies. Experimental results demonstrate substantial performance improvements with few-shot prompting, achieving macro-average F1 scores as high as 0.85. Our findings highlight the significant advantages of incorporating exemplar-based reasoning, underscoring that CoT and RAG approaches markedly enhance small ODLLMs' capabilities in accurately classifying complex network attacks under stringent resource constraints.

\end{abstract}
% \IEEEpeerreviewmaketitle
\section{Introduction}

The proliferation of Internet of Things (IoT) sensors in residential and industrial environments has significantly accelerated digital transformation by timely and effective processing of precise data for rapid decision-making \cite{iotddos2}. However, the extensive deployment of IoT devices also introduced network risks, especially with the increasing frequency and complexity of Distributed Denial of Service (DDoS) attacks and the introduction of more sophisticated and adaptive attack combinations \cite{iotddos1}. These evolving threats pose substantial challenges for conventional detection methodologies, underscoring the necessity for advanced detection mechanisms.

The recent emergence of Generative Artificial Intelligence (GenAI), particularly Large Language Models (LLMs), provides promising capabilities for enhancing cybersecurity by identifying complex and dynamic attack patterns in real-time \cite{llmabnormal}\cite{ylu1}. However, employing cloud-based LLM solutions introduces critical privacy concerns and latency issues, particularly pertinent in real-time threat detection scenarios and sensitive data environments \cite{oct1}. 
On the other hand, deploying intelligence at the edge with On-Device Large Language Models (ODLLMs) has been actively investigated \cite{llmqw1, oct2}. By enabling effective decision-making directly on edge devices, ODLLM-based solutions can alleviate privacy risks and latency constraints \cite{qunodllm} \cite{yl3}. Nonetheless, the constraints on token input length and the limited capabilities of intrinsic model understanding often result in suboptimal detection performance without adequate background knowledge and tailored support mechanisms.

Our previous work utilized a structured knowledge base to increase ODLLM-enabled detectors' performance, but the smaller model is still struggling when handling multiple attacks \cite{qunddos}.
Recent efforts in enhancing LLM reasoning capabilities have emphasized either structural prompting or compute-adaptive inference. Weng~\cite{weng2025thinking} frames LLM reasoning as a balance between fast heuristic-driven responses and slow deliberative reasoning. Techniques such as Chain-of-Thought (CoT) prompting and retrieval-augmented generation (RAG)  are presented as means to improving reliability on complex tasks \cite{wei2022chain} \cite{lewis2020retrieval}.
Complementary to this cognitive framing, Snell et al. study the trade-off between model size and test-time compute, showing that iterative revision and verifier-guided search can outperform naively scaling parameters \cite{snell2024scaling}.  However, how to integrate those methods to improve ODLLM detection performance still needs to be solved.

To address these challenges, this paper proposes an innovative detection framework integrating RAG and CoT methodologies to enhance ODLLM capabilities specifically for IoT network security. Our contributions are as follows:
(1) We perform an in-depth analysis of DDoS attack characteristics and their implications for ODLLM-based detection. (2) We introduce a novel CoT-guided inference mechanism tailored for resource-constrained edge models to improve reasoning and accuracy. (3) We compare and design a compact RAG algorithm that is optimized for small-model constraints to facilitate effective integration of structured knowledge bases. (4) We perform comprehensive experimental evaluations for different ODLLMs and validate the effectiveness and robustness of our proposed framework. It demonstrates that our enhanced ODLLM system achieves comparable detection accuracy to larger cloud-based models while maintaining the operational advantages of edge-based deployment.

The subsequent sections are organized as follows. The system model and problem formulation are presented in Section II. The proposed CoT-assisted RAG design is developed in Section III. The simulation results are presented in Section IV. Finally, Section V provides the concluding remarks for this paper.

% \begin{figure}
% \setlength{\abovedisplayskip}{3pt}
% 	\setlength{\belowdisplayskip}{3pt}
% \centering
% \includegraphics[width=0.90\linewidth]{AnonymousSubmission/LaTeX/system.jpg}
% \caption{System model.\label{sys}}
% \vspace{-0.5cm}
% \end{figure}	

% \section{Related work and Limitation of ODLLM}
% %这一章重点讨论之前观察到的odllm存在的限制，不如不懂基础的概念，数学推理很差，缺乏判断力，以及谷歌和openai的文章带来的启发，我们要做的提升的理论基础

% \subsection{Related Work}

\section{System Model and Problem Formulation}

\subsection{System Model}

\begin{figure}[h]
    \centering
    \includegraphics[width=0.8\columnwidth]{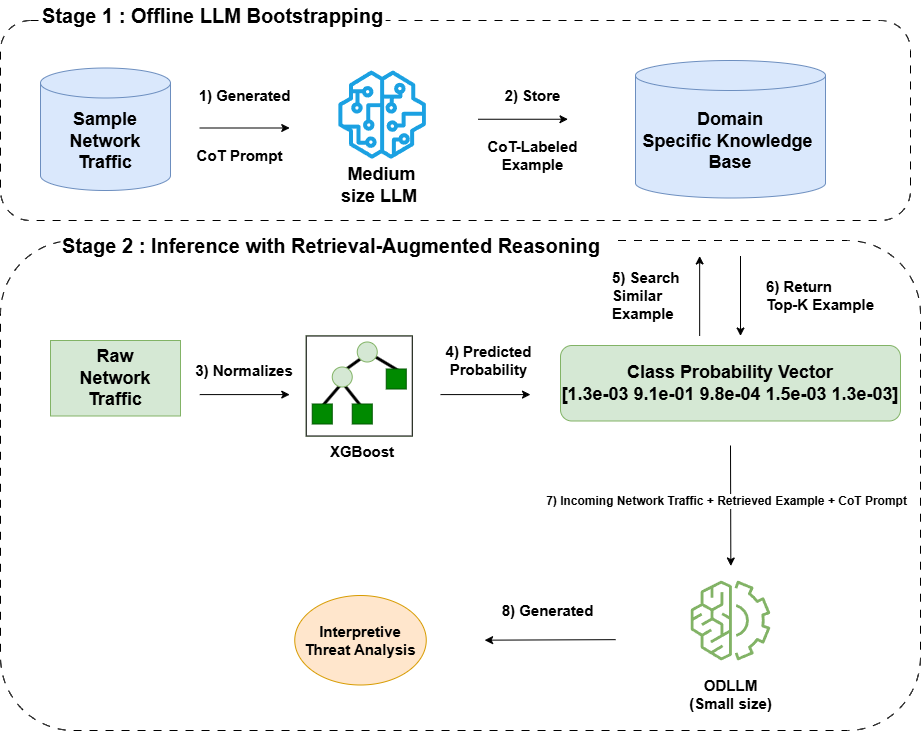} % Adjust width relative to the column
    \caption{Proposed System Pipeline.}
    \label{fig:pipe}
\end{figure}

As shown in Fig. \ref{fig:pipe}, our cybersecurity attack detection framework consists of two stages. 
In Stage 1, we utilize a larger-sized LLM as a teacher model to extract key insights from historical datasets of collected attacks offline, thereby generating step-by-step reasoning outputs based on CoT prompting and constructing a domain-specific knowledge base (KB). This will lay the foundation for small-sized ODLLM to perform its detection and overcome its input limitations and constraint reasoning capabilities. Unlike standard distillation, this process does not rely on soft labels. Instead, the teacher model provides detailed CoT-based reports for each sample, building new reasoning steps toward a well-formed conclusion. These serve as demonstrations for the small-sized student ODLLM model. ODLLM will learn to follow the reasoning process and independently arrive at similar conclusions.

Based on Stage 1, Stage 2 will perform inference with RAG-assisted ODLLM. Enlightened by paper \cite{snell2024scaling}, the raw network traffic will first be transformed into feature embeddings generated by a pre-trained XGBoost classifier. We then compute the Euclidean distance between samples and sort the results in ascending order. For few-shot prompting, we experimented with providing the ODLLM model with one to three example samples. Notably, the objective is to utilize ODLLM's inference time (test time compute) capabilities for slow, deliberate, and logically coherent step-by-step reasoning. By performing the RAG search with XGBoost output, the network traffic will be combined with retrieved examples and the CoT prompt, and then fed into ODLLM for attack detection. 

% \subsubsection{Stage 1: Offline LLM Bootstrapping}

% In the first stage, a medium-sized teacher model is used to generate step-by-step reasoning outputs based on CoT prompting. Unlike standard distillation, this process does not rely on soft labels. Instead, the teacher model provides detailed CoT-based reports for each sample, building new reasoning steps toward a well-formed conclusion. These serve as demonstrations for the small-sized student ODLLM model. ODLLM will learn to follow the reasoning process and independently arrive at similar conclusions.

% \subsubsection{Stage 2: Inference with RAG}
% In the second stage, we leverage XGBoost to retrieve relevant samples by pre-processing numerical data and predicting class probabilities. We then compute the Euclidean distance between samples and sort the results in ascending order. For few-shot prompting, we experimented with providing the ODLLM model with one to three example samples, as evaluated in Section~\ref{experience_setup}. Notably, the objective is to utilize ODLLM's inference time(test time compute) capabilities for slow, deliberate, and logically coherent step-by-step reasoning (system 2).

% Stage 1  bootstraps a domain–specific knowledge base (KB): a small set of canonical flows is fed to the ODLLM together with a
% three-step Chain-of-Thought (CoT) prompt; the resulting $\langle\!\text{network traffic},\text{label},\text{rationale}\!\rangle$ triples are stored in the KB for later retrieval. Stage~2 (\emph{online}) executes continuously on an edge gateway that fronts a swarm of IoT devices.

With important feature ranking from our previous research \cite{qunddos}, incoming packets are aggregated into unidirectional flows.
Each flow is encoded as a nine-dimensional vector corresponding to 9 top-ranking features:
\begin{equation}
  x \;=\;
   \bigl[
     \text{proto},\;
     r,\;
     \text{IAT},\;
     s,\;
     \alpha_{\text{PSH}},
     \alpha_{\text{ACK}},
     \alpha_{\text{SYN}},
     \alpha_{\text{RST}},
     \alpha_{\text{FIN}}
   \bigr]\in\mathbb R^{9},
  \label{eq:feature}
\end{equation}
where \textit{proto} is the IP-layer protocol number,
$r$ is the packet-rate (pps), \textit{IAT} the mean inter-arrival time
(ms), $s$ the average payload length (bytes) and the five $\alpha_{i}\in\{0,1\}, i\in\{\text{PSH},\text{ACK},\text{SYN},\text{RST}, \text{FIN}\}$
are TCP flags.

% \paragraph{Lightweight prior model.}
The feature vector is normalized and passed to an XGBoost classifier
$f_{\psi}:\mathbb R^{9}\!\rightarrow\!\Delta^{C}$ with $C$ attack families labels.  
It produces a softmax vector
\begin{equation}
  \mathbf p \;=\; f_{\psi}(x), \qquad
  \mathbf p\in\Delta^{C},
  \label{eq:prior}
\end{equation}
which acts both as a coarse prediction and as a compact semantic signature for retrieval.

% \paragraph{Nearest-neighbor retriever.}
Given $\mathbf p$, the retriever consults the KB $\mathcal M=\{(\mathbf p_i,x_i,y_i)\}_{i=1}^{N}$ and returns the indices of the $k$ most similar prototypes under Euclidean distance:
\begin{equation}
  R_k(\mathbf p) \;=\;
  \operatorname*{arg\,top\text{-}k}_{j\le N}
  \bigl\|\,\mathbf p-\mathbf p_j\,\bigr\|_2 ,
  \label{eq:retriever}
\end{equation}
where $x_i$ is the archived flow and $y_i$ is the ground-truth label of that flow.

% \paragraph{Prompt composer and LLM inference.}
The live flow description $\mathcal{D(x)}$, the retrieved exemplar block, and the fixed CoT template are concatenated into a prompt presented to ODLLM $g_{\theta}$.  
With decoding temperature $T\!=\!0$ the model returns a deterministic label $\hat {y}$ and a natural-language rationale $\pi$:
\begin{equation}
  (\hat y,\pi) \;=\;
  g_{\theta}\!\bigl(
     P_{\eta}\bigl(R_k(\mathbf p),\;\mathcal D(x)\bigr)
  \bigr),
  \label{eq:llm}
\end{equation}
where $P_{\eta}$ denotes the template-driven prompt builder.

\subsection{Problem Formulation}
Let the streaming data be
$\mathcal D=\{(x_t,y_t)\}_{t\ge 1}$ with $x_t\in\mathbb R^{9}$ and ground-truth $y_t\in\mathcal Y=\{1,\dots,C\}$. Define the composite detector
\begin{equation}
  h_{\Phi}(x)
  \;=\;
  g_{\theta}\!\bigl(
    P_{\eta}\bigl(
      R_k\bigl(f_{\psi}(x)\bigr),
      \;\mathcal D(x)
    \bigr)
  \bigr),
  \label{eq:detector}
\end{equation}
with the parameter set $\Phi=\{\theta,\psi,\eta,\mathcal M\}$.

The design objective is to maximize the long-run detection accuracy
\begin{equation}
  \Phi^{\star}
  \;=\;
  \arg\max_{\Phi}\;
  \mathbb E_{(x,y)\sim\mathcal D}
  \bigl[\mathbf 1\{h_{\Phi}(x)=y\}\bigr],
  \label{eq:objective}
\end{equation}
By setting $k=0$,  the equation \eqref{eq:detector} reduces to a pure CoT baseline without retrieval.
In the next section, we will solve the above problem via CoT design and RAG enhancement.

\section{ODLLM Detector Enhancement with CoT Reasoning}

\subsection{Chain-of-Thought Prompting}
Chain-of-thought (CoT) prompting enhances the reasoning capabilities of large language models (LLMs) by encouraging them to generate intermediate reasoning steps before producing a final answer. Rather than jumping directly to conclusions, the model is guided to break down the task into smaller, logically connected steps. 
% For instance, when solving a math problem like: “After Jane gives 2 flowers to her mom she has 10... then after she gives 3 to her dad she will have 7...”, the model follows the reasoning process to arrive at the final answer of 7 .
This structured decomposition mirrors the way cybersecurity professionals approach complex incident analysis \cite{wei2022chain}. 
 For example, when investigating a DDoS attack, analysts typically examine traffic flows, rate limits, source IP distributions, and attack durations step-by-step to isolate the root cause and determine the appropriate response. Similarly, CoT prompting enables LLMs to build a line of reasoning that mimics this analytical process.
In this paper, we implement a rule-based prompt structure that elicits domain-specific CoT reasoning for cybersecurity reporting. We apply the “Let's think step by step" prefix strategy as demonstrated by \cite{kojima2022large}.

\subsection{Chain‐of‐Thought (CoT) Baseline}
\label{sec:cot}

% NEED UPDATE PROMPT     
% \begin{figure}[h]
% \setlength{\abovedisplayskip}{3pt}
% 	\setlength{\belowdisplayskip}{3pt}
%     \centering
%     \includegraphics[width=0.8\columnwidth]{figures/CoT_fig.png} % Adjust width relative to the column
%     \caption{Chain-of-thought prompt template}
%     \label{fig:cot_temp}
% \end{figure}

Our starting point is a rule-guided CoT prompt that allows ODLLM to emulate the deterministic logic traditionally encoded in signature-based intrusion systems.  
% Figure~\ref{fig:cot_temp} shows aconcrete example.  
The prompt is organized into three consecutive blocks:

\begin{enumerate}
  \item \textbf{Data description:} A natural-language rendering
        $\mathcal D(x)$ of the nine-dimensional feature vector
        in~\eqref{eq:feature}.
  \item \textbf{Instruction:} a one-shot classification directive that (i) constrains the output domain to the five DDoS flood families and (ii) forces the model to emit the final answer exactly once, in the pattern \texttt{``The~answer~is <LABEL>''}.
  \item \textbf{Knowledge base (KB):} A three-step reasoning scaffold
        $\mathcal T_{\text{CoT}}$ distilled from traffic-engineering
        heuristics:
        \begin{itemize}
       
            \item \textit{Packet-size \& rate gate.}
                  Reject benign flows whose mean packet length is
                  $>60$ B or whose packet rate is $<1$ pps
                  (empirically determined thresholds).
            \item \textit{Protocol branch.}
                  Distinguish ICMP and UDP floods
                  directly via the \texttt{proto} field.
            \item \textit{TCP flag analysis.}
                  If the protocol is TCP, map the tuple $(\alpha_{\text{PSH}},\alpha_{\text{ACK}},\alpha_{\text{RST}}, \alpha_{\text{FIN}})$ to \textsc{PSH/ACK}, \textsc{RST/FIN} or                   fallback to \textsc{TCP} flood.
         \end{itemize}
\end{enumerate}

\vspace{0.3em}
\noindent
The combined prompt is therefore $\text{Prompt}(x)= \mathcal D(x)\,\Vert\,\text{Instruction}\,\Vert\,
\mathcal T_{\text{CoT}}$, where ``$\Vert$'' denotes string concatenation.

 \section{RAG enhanced ODLLM Detection Design}

\subsection{Backbone of RAG }

The effectiveness of Retrieval Augmented Generation (RAG) depends strongly on the design of the retriever algorithm. In our work, we evaluate multiple retriever models, including BERT-based BGE (BAAI General Embedding), a Multi-Layer Perceptron (MLP), and XGBoost, to determine which best identifies relevant documents for each query.

\begin{figure*}[t!]
    \centering
    % Subfigure (a)
    \begin{subfigure}[b]{0.27\textwidth}
        \centering
        \includegraphics[width=\linewidth]{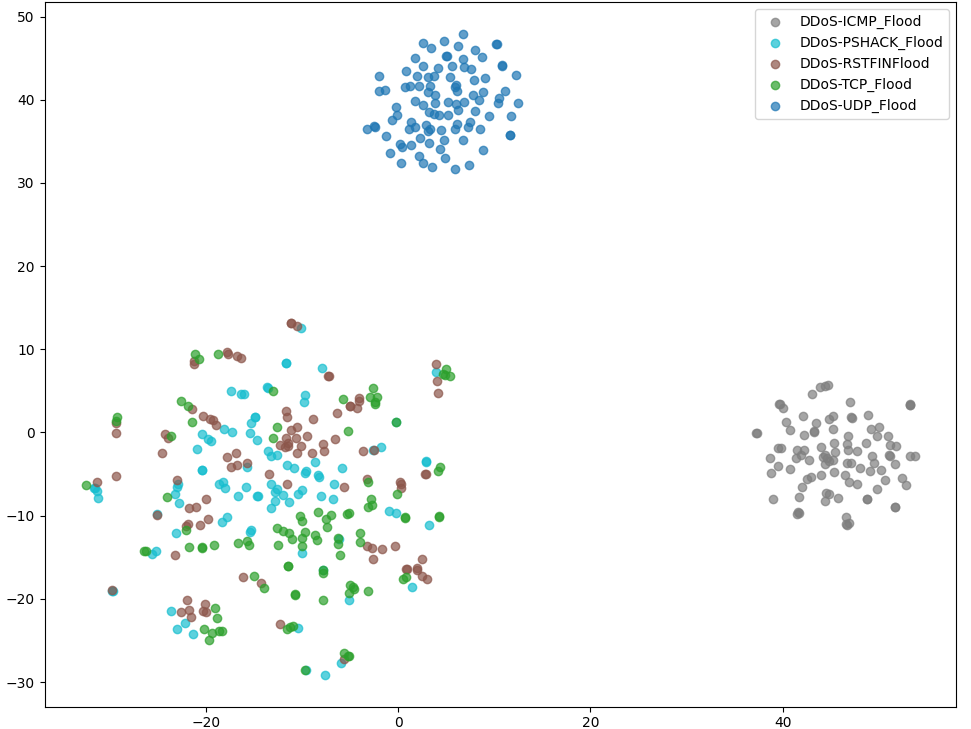}
        \caption{BGE base Embeddings}
        \label{fig:bge_SNE}
    \end{subfigure}
    \hfill % This command adds horizontal space between the figures
    % Subfigure (b)
    \begin{subfigure}[b]{0.27\textwidth}
        \centering
        \includegraphics[width=\linewidth]{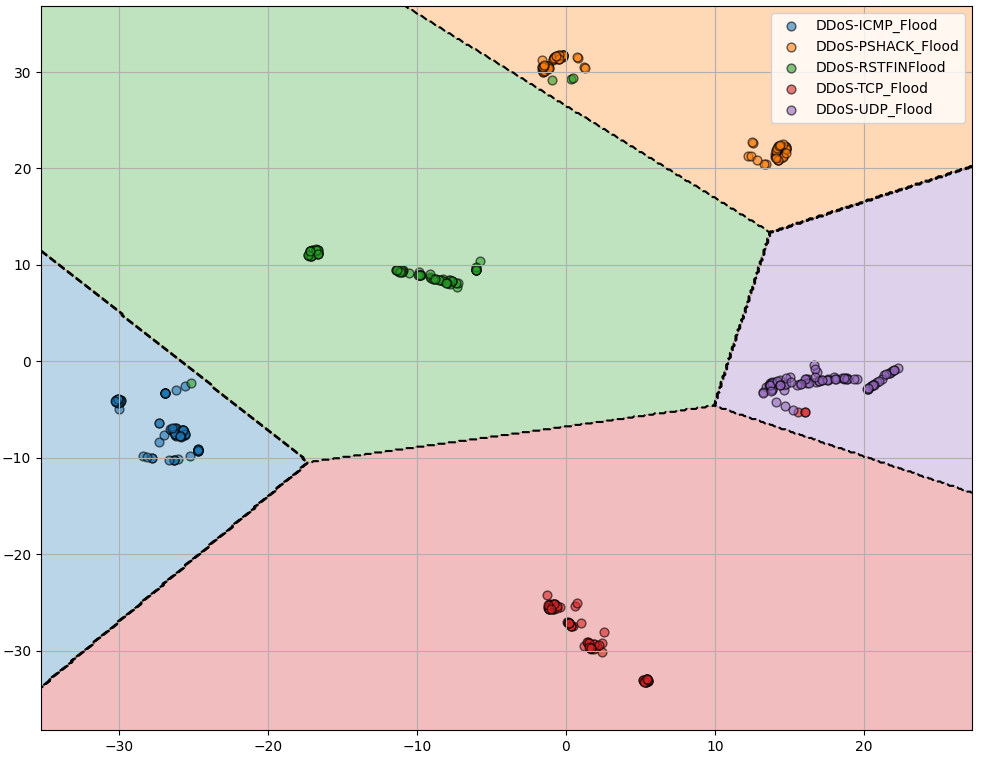}
        \caption{MLP base Embeddings}
        \label{fig:MLP_SNE}
    \end{subfigure}
    \hfill % This command adds horizontal space between the figures
    % Subfigure (c)
    \begin{subfigure}[b]{0.32\textwidth}
        \centering
        \includegraphics[width=\linewidth]{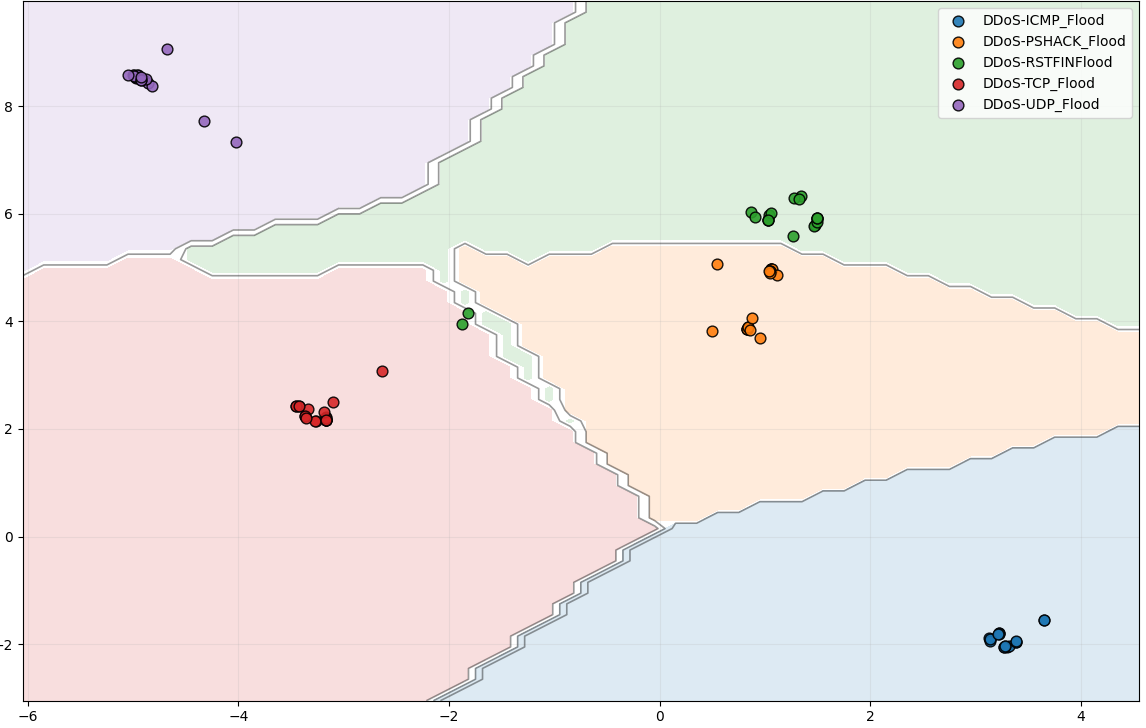}
        \caption{XGBoost base Embeddings}
        \label{fig:XGB_SNE}
    \end{subfigure}
    % This is the main caption for the entire figure
    \caption{2D t-SNE projections of network traffic embeddings from three different base models.}
    \label{fig:all_sne_projections}
\end{figure*}

\textbf{BGE Base.}
Our first retriever model is the \textit{bge-base-en-v1.5}, a fine-tuned BERT-base model from Hugging Face, to generate 768-dimensional sentence embeddings. This model is tailored for semantic similarity, retrieval, and clustering. However, after projecting these embeddings using t-SNE into 2D space (Figure~\ref{fig:bge_SNE}), we observe that BGE fails to clearly separate subcategories of TCP flood attacks (e.g., PSH/ACK and RST/FIN). This indicates that our network traffic data is not linearly separable in this embedding space, motivating the exploration of alternative approaches.

\textbf{MLP-based Embedding.}
To investigate whether a task-specific embedding model can improve separability, we design and train a Multi-Layer Perceptron (MLP) classifier. The MLP also takes network traffic data from our knowledge database and embeds it into a 16-dimensional latent space using a series of linear layers with ReLU activations and early stopping based on validation loss to avoid overfitting. The classifier is trained using cross-entropy loss with label smoothing on stratified splits of the data to ensure balanced class representation. 
After training, we extract the 16-dimensional embeddings and project them using t-SNE into 2D space for visualization. As shown in Figure~\ref{fig:MLP_SNE}, we observe a better separation between different categories of attacks, unlike the general-purpose BGE model. However, some regions still exhibit overlapping clusters or include irrelevant types, suggesting room for further improvement.

\textbf{XGBoost}
Recognizing the limitations of purely embedding-based approaches, we also investigate the use of XGBoost, a gradient boosting framework well-suited for tabular data ~\cite{grinsztajn2022tree}.
Unlike embedding models, XGBoost directly builds an ensemble of decision trees that iteratively correct previous errors, which better capture complex patterns in network traffic features. While XGBoost is primarily a classifier, we treat the standardized 9 numerical input features as fixed embeddings and apply t-SNE to project them into a 2D space. As shown in Figure~\ref{fig:XGB_SNE}, the resulting layout reveals that the non-linear decision boundaries can already capture the differences between attack categories.
For the training phase, we use the extracted numerical features from network traffic data. Then the model is configured for multi-class classification with the objective function \texttt{multi:softmax}, optimized to distinguish among the attack categories. Key hyper-parameters include a maximum tree depth of 6, a learning rate of 0.1, and 100 boosting iterations (estimators). This setup balances model complexity and overfitting risk while enabling efficient training on moderate-sized datasets.

% Compared to the MLP, XGBoost offers several practical advantages. First, it requires substantially less training time and computational resources because it does not rely on gradient backpropagation over multiple layers but instead incrementally builds decision trees. This makes it more cost-efficient, particularly on limited IoT devices. Second, XGBoost often achieves competitive or better accuracy on tabular data without extensive hyperparameter tuning or the risk of overfitting that deep neural networks sometimes face, especially on relatively small or imbalanced datasets.
% In summary, XGBoost serves as a powerful, resource-efficient retriever that complements MLP embeddings by offering robust predictive accuracy with lower computational cost, making it a pragmatic choice for retrieval tasks in our RAG system.

% ============================================

\subsection{Probability-Guided Retrieval–Augmented Generation (Prob-RAG)}
\label{sec:probrag}

XGBoost constructs additive ensemble models consisting of decision trees, optimized in a forward stage-wise manner. The training procedure minimizes the following regularized objective function:
\begin{equation}
\setlength{\abovedisplayskip}{3pt}
	\setlength{\belowdisplayskip}{3pt}
  \mathcal{L}^{(t)} = \sum_{i=1}^{n} l\bigl(y_i,\hat{y}_i^{(t-1)}+f_t(x_i)\bigr) + \Omega(f_t),
\end{equation}
where \( l(\cdot) \) is a differentiable convex loss function, such as logistic or multinomial log-loss, \(y_i\) is the ground-truth label, and \(\hat{y}_i^{(t-1)}\) is the prediction from the previous iteration.

The regularization term \(\Omega(f)\) controls the complexity of the tree model to prevent overfitting, defined as:
\begin{equation}
\setlength{\abovedisplayskip}{3pt}
	\setlength{\belowdisplayskip}{3pt}
  \Omega(f) = \gamma T + \frac{1}{2}\lambda\|w\|^2,
\end{equation}
where \(T\) denotes the number of leaves in the tree, \(w\) represents the vector of leaf weights, and \(\gamma, \lambda\) are regularization hyperparameters. Specifically, \(\gamma\) penalizes tree complexity by restricting the number of leaves, while \(\lambda\) provides \(L_2\)-regularization on leaf weights.

% \paragraph{Feature Embedding and Probability Signature.}
We represent each network flow as a nine-dimensional numeric feature vector capturing protocol type, packet rate, inter-arrival time, packet size, and TCP flags. Each vector is standardised (zero mean, unit variance) and used directly as the embedding signature for retrieval. Additionally, an XGBoost model $f_{\psi}$ trained to classify DDoS types outputs a class-probability vector $\mathbf{p}=f_{\psi}(x)\in\Delta^{5}$, providing both a coarse-grained prediction and an alternative semantic representation for retrieval.

% \paragraph{Nearest Neighbour Retrieval in Feature Space.}
The exemplar database $\mathcal{M}_P$ stores the standardised feature vectors of previously observed and labelled flows:
\begin{equation}     
\setlength{\abovedisplayskip}{3pt}
	\setlength{\belowdisplayskip}{3pt}
\mathcal{M}_P = \{(x_i, \mathcal D(x_i), y_i)\}_{i=1}^{N},
 \end{equation}
where $x_i$ is the normalised numeric feature vector, $\mathcal D(x_i)$ is the textual description, and $y_i$ is the corresponding label.

At inference time, we retrieve the top-$k$ exemplars closest to the current flow’s normalised vector $x$ using Euclidean distance:
\begin{equation}    
\setlength{\abovedisplayskip}{3pt}
	\setlength{\belowdisplayskip}{3pt}
R_k^P(x) = \operatorname*{arg\,top\text{-}k}_{j}\|x - x_j\|_2.
\end{equation}

% \paragraph{Prompt Construction and LLM Inference.}
Retrieved exemplars are concatenated with the live flow description and a fixed Chain-of-Thought (CoT) instruction template into a compact prompt:
\begin{equation}      
\setlength{\abovedisplayskip}{3pt}
	\setlength{\belowdisplayskip}{3pt}
\text{Prompt}_{\text{Prob}} = \langle E_{R_k^P(x)}\rangle \;\|\; \mathcal T_{\text{CoT}} \;\|\; \mathcal D(x).
\end{equation}
This prompt is submitted to a local small language model to yield a final classification.

%--------------------------------------------------------------------------------
% \begin{figure}[h]
% \setlength{\abovedisplayskip}{3pt}
% 	\setlength{\belowdisplayskip}{3pt}
%     \centering
%     \includegraphics[width=0.8\columnwidth]{figures/Figure_2D.png} % Adjust width relative to the column
%     \caption{2D t-SNE Projection of Network Traffic BGE base Embeddings}
%     \label{fig:bge_SNE}
% \end{figure}

% %--------------------------------------------------------------------------------

% \begin{figure}[h]
% \setlength{\abovedisplayskip}{3pt}
% 	\setlength{\belowdisplayskip}{3pt}
%     \centering
%     \includegraphics[width=0.8\columnwidth]{figures/2d_mlp.png} % Adjust width relative to the column
%     \caption{2D t-SNE Projection of Network Traffic MLP base Embeddings}
%     \label{fig:MLP_SNE}
% \end{figure}

% %--------------------------------------------------------------------------------

% \begin{figure}[h]
% \setlength{\abovedisplayskip}{3pt}
% 	\setlength{\belowdisplayskip}{3pt}
%     \centering
%     \includegraphics[width=0.8\columnwidth]{figures/2d_xg.png} % Adjust width relative to the column
%     \caption{2D t-SNE Projection of Network Traffic XGBoost base Embeddings}
%     \label{fig:XGB_SNE}
% \end{figure}

%--------------------------------------------------------------------------------

\begin{table*}[ht]
\centering
\begin{tabular}{lllllllll}
\hline
                              & Attack   & ICMP & UDP  & TCP  & PSH/ACK & RST/FIN & Benign & Macro avg \\ \hline
\multirow{5}{*}{Gemma3 1B}    & No KB    & 0.42 & 1    & 0.49 & 0.33    & 0       & 0      & 0.37      \\
                              & Short KB & 0.99 & 0.65 & 0.28 & 0       & 0       & 0      & 0.32      \\
                              & COT      & 0.2  & 0.62 & 0.31 & 0.3     & 1       & 0.13   & 0.43      \\
                              & One-Shot & 0.77 & 0.98 & 0.35 & 0       & 0.25    & 0.74   & 0.51      \\
                              & Few-Shot & 0.96 & 0.91 & 0.3  & 0.18    & 1       & 0.8    & 0.69      \\ \hline
\multirow{5}{*}{Gemma3 4B}    & No KB    & 0    & 1    & 0.12 & 0.82    & 0.98    & 0.08   & 0.5       \\
                              & Short KB & 1    & 1    & 0.33 & 1       & 0.98    & 0      & 0.72      \\
                              & COT      & 0    & 0.5  & 0.24 & 0.83    & 0.95    & 0.4    & 0.49      \\
                              & One-Shot & 1    & 0.95 & 0.68 & 0.95    & 0.98    & 0.8    & 0.89      \\
                              & Few-Shot & 1    & 0.97 & 0.47 & 0.86    & 0.98    & 0.58   & 0.81      \\ \hline
\multirow{5}{*}{Llama 3.2 1B} & No KB    & 0.22 & 0.72 & 0.2  & 0       & 0       & 0.38   & 0.25      \\
                              & Short KB & 0.28 & 0.86 & 0.21 & 0       & 0       & 0.22   & 0.26      \\
                              & COT      & 0.49 & 0.55 & 0.2  & 0.19    & 0.29    & 0.15   & 0.31      \\
                              & One-Shot & 0.97 & 0.99 & 0.29 & 0       & 0       & 0.85   & 0.52      \\
                              & Few-Shot & 0.94 & 0.98 & 0.3  & 0       & 0       & 0.95   & 0.53      \\ \hline
\multirow{5}{*}{Llama 3.2 3B} & No KB    & 0    & 1    & 0.02 & 1       & 1       & 0.21   & 0.54      \\
                              & Short KB & 1    & 1    & 0.25 & 1       & 0       & 0.76   & 0.67      \\
                              & COT      & 0.91 & 0.97 & 0.34 & 0.89    & 0.99    & 0.26   & 0.73      \\
                              & One-Shot & 0.91 & 0.89 & 0.42 & 0.93    & 0.98    & 0.69   & 0.8       \\
                              & Few-Shot & 0.91 & 0.92 & 0.52 & 0.94    & 0.99    & 0.57   & 0.81      \\ \hline
\end{tabular}
\caption{Precision Score of LLMs on DDoS Attack Types Using Different Reasoning Methods.\label{recall}}
\end{table*}

\section{Experiment Performance Evaluation}
%仿真思路：用之前的文章做benchmark，重点强调新的改进对模型的提升，以及一些观察到的现象。分类别讨论。

% \subsection{Dataset and Experiment Setting}

We use Ollama to retrieve ODLLMs with our constructed KB on Desktop with NVIDIA RTX 4090 and Intel I9-13900KF \cite{ollama}. 
% We test our model's performance with the CICIoT 2023 dataset \cite{ciciot2023}. Our source code is released on GitHub (https://github.com/claudwq/Intelligent-IoT-Attack-Detection-Design-via-LLM-with-Feature-Ranking-Based-Knowledge-Base-Design.git).
We consider the latest small-sized models as follows:
\textbf{Llama 3.2 3B \& 1B} is a compact variant in the Llama 3 series with 3 billion parameters and 1 billion parameters, optimized for multilingual tasks and large-scale text processing \cite{llama3}. 
\textbf{Gemma 3 4B \& 1B} Gemma is a lightweight, family of models from Google built on Gemini technology. The Gemma 3 models are multimodal—processing text and images—and feature a 128K context window with support for over 140 languages \cite{gemma3}.
To demonstrate the effectiveness of our proposed design, we consider the KB enhanced method from our previous research \cite{qunddos} as a benchmark.

To evaluate our proposed intelligent IoT attack detection design, we consider the dataset CICIOT 2023, which is a real-time dataset and benchmark for large-scale attacks on evaluating intrusion detection systems in IoT environments, featuring diverse network traffic types, including normal and malicious behaviors, across multiple IoT protocols and attack scenarios \cite{ciciot2023}. We chose 500 high-quality samples each in 5 different types of DDoS attacks and Benign traffic.
\textbf{ICMP Flood Attack} overwhelms the target with a high volume of ICMP echo requests, causing the network to become congested and unresponsive.\textbf{UDP Flood Attack} sends a large number of UDP packets to random ports on the target server, forcing it to process unnecessary requests.
\textbf{TCP SYN Flood Attack} exploits the TCP handshake mechanism by sending numerous SYN packets without completing the handshake, consuming server resources.
\textbf{TCP PSH+ACK Flood Attack} sends a large number of TCP packets with the PSH (Push) and ACK (Acknowledgment) flags set to overwhelm the target's processing capabilities. 
Notably, \textbf{PSHAKC} and \textbf{RSTFIN} attack type is a subset of \textbf{DDOS TCP attack}, which requires the model to reason deeper and avoid just mimicking the response. For instance, when only providing a knowledge base to ODLLM, it might just give a short answer once it finds a DDoS TCP attack and may not investigate further.

% \subsection{Types of DDoS Attack}

% We consider four types of DDoS attacks and their characteristics:

% \textbf{ICMP Flood Attack} overwhelms the target with a high volume of ICMP echo requests, causing the network to become congested and unresponsive.
% This attack usually causes increased bandwidth consumption, degraded service performance, and potential downtime.

% \textbf{UDP Flood Attack} sends a large number of UDP packets to random ports on the target server, forcing it to process unnecessary requests.
% It usually has random destination ports and stateless protocol exploitation.
% UDP flood will increase CPU usage and deny legitimate service requests.

% \textbf{TCP SYN Flood Attack} exploits the TCP handshake mechanism by sending numerous SYN packets without completing the handshake, consuming server resources. It usually has elevated SYN flags, half-open TCP connections, and spoofed IP addresses. It will exhaust connection tables, leading to an inability to establish new legitimate connections.

% \textbf{TCP PSH+ACK Flood Attack} sends a large number of TCP packets with the PSH (Push) and ACK (Acknowledgment) flags set to overwhelm the target's processing capabilities. It usually involves high volumes of PSH and ACK packets that mimic normal traffic patterns, making them difficult to filter. This flood increases processing overhead, leads to resource depletion, and can cause potential service crashes.

% \subsection{Evaluation Metrics}

To evaluate the performance of our attack detection framework, we adopt the macro-average F1 score as the primary metric. Since our classification task involves six balanced classes with 500 samples each, the macro-average F1 score ensures equal weighting across all classes, providing a fair assessment of the model's overall performance.
The F1 score is the harmonic mean of precision and recall, capturing the balance between false positives and false negatives:
$\text{F1} = 2 \times \frac{\text{Precision} \times \text{Recall}}{\text{Precision} + \text{Recall}}$.
Where \textbf{Precision} measures how many of the predicted positive instances are actually correct. \textbf{Recall} measures how many of the actual positive instances were correctly identified.
To compute the macro-average F1 score, the F1 score is calculated independently for each class and then averaged:
$\text{Macro-F1} = \frac{1}{N} \sum_{i=1}^{N} \text{F1}_{i}$,
where \( N \) is the number of classes. This metric is particularly appropriate when classes are evenly distributed, as it avoids bias toward any specific class and provides a holistic view of model performance across all types of network traffic.

\begin{table*}[ht]
\centering
\begin{tabular}{cclllllll}
\hline
\multicolumn{1}{l}{\cellcolor[HTML]{FFFFFF}}           & \multicolumn{1}{l}{Attack}       & \multicolumn{1}{c}{ICMP}     & \multicolumn{1}{c}{UDP}      & \multicolumn{1}{c}{TCP}      & \multicolumn{1}{c}{PSH/ACK}  & \multicolumn{1}{c}{RST/FIN}  & \multicolumn{1}{c}{Benign} & \multicolumn{1}{c}{Macro avg} \\ \hline
\cellcolor[HTML]{FFFFFF}                               & No KB                            & 0.59                         & 0.42                         & 0.57                         & 0.44                         & 0                            & 0                          & 0.34                           \\
\cellcolor[HTML]{FFFFFF}                               & Short KB                         & 0.96                         & 0.77                         & 0.43                         & 0                            & 0                            & 0                          & 0.36                           \\
\cellcolor[HTML]{FFFFFF}                               & COT                              & 0.31                         & 0.24                         & 0.09                         & 0.05                         & 0.03                         & 0.06                       & 0.13                           \\
\cellcolor[HTML]{FFFFFF}                               & One-Shot                         & 0.73                         & 0.83                         & 0.34                         & 0                            & 0                            & 0.59                       & 0.42                           \\
\multirow{-5}{*}{\cellcolor[HTML]{FFFFFF}Gemma3 1B}    & Few-Shot                         & 0.85                         & 0.8                          & 0.38                         & 0.01                         & 0                            & 0.82                       & \textbf{0.48}                  \\ \hline
\cellcolor[HTML]{FFFFFF}                               & No KB                            & 0                            & 0.98                         & 0.11                         & 0.85                         & 0.99                         & 0.1                        & 0.51                           \\
\cellcolor[HTML]{FFFFFF}                               & Short KB                         & 1                            & 0.98                         & 0.5                          & 0                            & 0.99                         & 0                          & 0.58                           \\
\cellcolor[HTML]{FFFFFF}                               & COT                              & 0                            & 0                            & 0.03                         & 0.11                         & 0.14                         & 0.01                       & 0.05                           \\
\cellcolor[HTML]{FFFFFF}                               & One-Shot                         & 0.83                         & 0.94                         & 0.78                         & 0.96                         & 0.98                         & 0.64                       & \textbf{0.85}                  \\
\multirow{-5}{*}{\cellcolor[HTML]{FFFFFF}Gemma3 4B}    & Few-Shot                         & 0.67                         & 0.94                         & 0.56                         & 0.91                         & 0.96                         & 0.57                       & 0.77                           \\ \hline
\cellcolor[HTML]{FFFFFF}                               & No KB                            & 0.19                         & 0.05                         & 0.04                         & 0                            & 0                            & 0.03                       & 0.05                           \\
\cellcolor[HTML]{FFFFFF}                               & \cellcolor[HTML]{FFFFFF}Short KB & \cellcolor[HTML]{FFFFFF}0.36 & \cellcolor[HTML]{FFFFFF}0.16 & \cellcolor[HTML]{FFFFFF}0.22 & \cellcolor[HTML]{FFFFFF}0    & \cellcolor[HTML]{FFFFFF}0    & 0.04                       & 0.13                           \\
\cellcolor[HTML]{FFFFFF}                               & \cellcolor[HTML]{FFFFFF}COT      & \cellcolor[HTML]{FFFFFF}0.27 & \cellcolor[HTML]{FFFFFF}0.25 & \cellcolor[HTML]{FFFFFF}0.22 & \cellcolor[HTML]{FFFFFF}0.04 & \cellcolor[HTML]{FFFFFF}0.02 & 0.05                       & 0.14                           \\
\cellcolor[HTML]{FFFFFF}                               & \cellcolor[HTML]{FFFFFF}One-Shot & \cellcolor[HTML]{FFFFFF}0.83 & \cellcolor[HTML]{FFFFFF}0.82 & \cellcolor[HTML]{FFFFFF}0.45 & \cellcolor[HTML]{FFFFFF}0    & \cellcolor[HTML]{FFFFFF}0    & 0.58                       & \textbf{0.45}                  \\
\multirow{-5}{*}{\cellcolor[HTML]{FFFFFF}Llama 3.2 1B} & \cellcolor[HTML]{FFFFFF}Few-Shot & \cellcolor[HTML]{FFFFFF}0.92 & \cellcolor[HTML]{FFFFFF}0.89 & \cellcolor[HTML]{FFFFFF}0.45 & \cellcolor[HTML]{FFFFFF}0    & \cellcolor[HTML]{FFFFFF}0    & 0.77                       & 0.5                            \\ \hline
\cellcolor[HTML]{FFFFFF}                               & \cellcolor[HTML]{FFFFFF}No KB    & \cellcolor[HTML]{FFFFFF}0    & \cellcolor[HTML]{FFFFFF}0.63 & \cellcolor[HTML]{FFFFFF}0.02 & \cellcolor[HTML]{FFFFFF}0    & \cellcolor[HTML]{FFFFFF}0.69 & 0.34                       & 0.28                           \\
\cellcolor[HTML]{FFFFFF}                               & \cellcolor[HTML]{FFFFFF}Short KB & \cellcolor[HTML]{FFFFFF}0.99 & \cellcolor[HTML]{FFFFFF}0.89 & \cellcolor[HTML]{FFFFFF}0.39 & \cellcolor[HTML]{FFFFFF}0.15 & \cellcolor[HTML]{FFFFFF}0    & 0.11                       & 0.42                           \\
\cellcolor[HTML]{FFFFFF}                               & \cellcolor[HTML]{FFFFFF}COT      & \cellcolor[HTML]{FFFFFF}0.8  & \cellcolor[HTML]{FFFFFF}0.82 & \cellcolor[HTML]{FFFFFF}0.47 & \cellcolor[HTML]{FFFFFF}0.68 & \cellcolor[HTML]{FFFFFF}0.75 & 0.17                       & 0.62                           \\
\cellcolor[HTML]{FFFFFF}                               & \cellcolor[HTML]{FFFFFF}One-Shot & \cellcolor[HTML]{FFFFFF}0.7  & \cellcolor[HTML]{FFFFFF}0.73 & \cellcolor[HTML]{FFFFFF}0.57 & \cellcolor[HTML]{FFFFFF}0.84 & \cellcolor[HTML]{FFFFFF}0.82 & 0.6                        & 0.71                           \\
\multirow{-5}{*}{\cellcolor[HTML]{FFFFFF}Llama 3.2 3B} & \cellcolor[HTML]{FFFFFF}Few-Shot & \cellcolor[HTML]{FFFFFF}0.83 & \cellcolor[HTML]{FFFFFF}0.79 & \cellcolor[HTML]{FFFFFF}0.62 & \cellcolor[HTML]{FFFFFF}0.85 & \cellcolor[HTML]{FFFFFF}0.79 & 0.6                        & \textbf{0.75}                  \\ \hline
                                                       &                                  & \multicolumn{1}{c}{}         & \multicolumn{1}{c}{}         & \multicolumn{1}{c}{}         & \multicolumn{1}{c}{}         & \multicolumn{1}{c}{}         & \multicolumn{1}{c}{}       &                               
\end{tabular}
\caption{F1 Score of LLMs on DDoS Attack Types Using Different Reasoning Methods.\label{f1}}
\end{table*}

The experimental results, summarized in Tables I and II, reveal that model performance is critically dependent on the chosen prompting strategy. We observe a distinct hierarchy of effectiveness: Few-Shot RAG significantly outperforms One-Shot RAG, which in turn is superior to knowledge base-enhanced (Short KB), CoT, and baseline Zero-Shot (No KB) methods. This finding holds true across both Gemma3 and Llama 3.2 model families. In baseline zero-shot configurations, all models struggled to process the traffic data, leading to a state of "mode collapse". This was particularly severe for smaller models; for example, Llama 3.2 1B (No KB) registered a macro average F1-score of just 0.05, demonstrating a near-total inability to differentiate between attack classes.

Our analysis highlights a contrast between the failure of abstract reasoning and the success of in-context learning.
The Inefficacy of CoT: The CoT strategy, intended to elicit step-by-step reasoning, proved ineffective and often detrimental. For instance, the Gemma3 1B with CoT experiment yielded a macro average F1-score of only 0.13, one of the lowest in our evaluation. Qualitative analysis of the model's outputs confirmed that the generated "reasoning" was frequently filled with logical fallacies, such as "protocol blindness" and misinterpretation of numerical values, constituting a form of confident hallucination. This indicates that forcing a model to "think" about a task it fundamentally does not understand leads to fabricated logic rather than genuine insight.

The introduction of just one (One-Shot) and three (Few-Shot) solved examples via RAG was the definitive factor in unlocking high-performance detection. This approach provides the model with a imitable template for analysis. The impact was dramatic: the macro F1-score for Llama 3.2 1B surged from 0.05 (No KB) to 0.50 (Few-Shot). Even smaller models learned to achieve high accuracy on specific classes, such as Llama 3.2 1B reaching 0.92 macro F1-score on ICMP floods with a few-shot prompt. This demonstrates that for structured tasks, LLMs excel not at abstract reasoning, but at analogical mapping from clear exemplars.

Model scale acts as a crucial amplifier for the effectiveness of a given prompting strategy. While a better strategy improves all models, a larger model is better equipped to capitalize on a sophisticated strategy.
The benefit of scale is most pronounced in the Few-Shot RAG setting, where larger models are more adept at "meta-learning" by abstracting a general methodology from examples. The performance gap between small and large models is widest here. The Gemma3 4B model (Few-Shot) achieved an impressive macro F1-score of 0.77, substantially outperforming its 1B counterpart (0.48). The most robust performance was delivered by the Llama 3.2 3B model under a few-shot regimen, which achieved the highest macro average F1-score of 0.75 across its family, with consistently high F1-scores on complex, nested classes like PSH/ACK (0.85) and RST/FIN (0.79). This underscores the synergistic effect of sufficient model scale and high-quality, example-based prompting.

In conclusion, our results provide strong evidence that while off-the-shelf ODLLMs are ill-suited for DDoS detection, their performance can be dramatically enhanced through carefully engineered, example-driven RAG prompts. The combination of a sufficiently scaled model (3B+ parameters) and a few-shot learning strategy creates a powerful and accurate system for classifying sophisticated network attacks in resource-constrained environments.

\section{Conclusions}

In this study, we presented a novel approach to enhancing the capabilities of small-scale ODLLMs for accurate and efficient detection of IoT-based DDoS attacks through CoT prompting and RAG.  Our work evaluates compact models like LLaMA 3.2 and Gemma 3 across a spectrum of prompting strategies. We found that abstract reasoning approaches, particularly standalone CoT prompting, fail spectacularly and suffer from severe logical fallacies.
In contrast, the RAG framework providing a few concrete examples proved transformative. This approach effectively re-purposes the ODLLM from a failed analyst into a highly efficient and accurate pattern-matching engine. This underscores that for ODLLMs operating on structured data, contextual grounding through relevant exemplars is not just beneficial—it is essential for achieving reliable performance.
This paradigm shift from abstract reasoning to guided, example-based mapping provides a robust and replicable methodology for deploying ODLLMs in critical, structured-data environments like network security. 
% Future research should build upon this foundation by exploring optimized retrieval techniques tailored to the dynamic nature of IoT attack patterns and investigating energy-aware implementations suitable for resource-limited edge devices.

	%\begin{thebibliography}{}
   \addtolength{\topmargin}{0.3in}
	\bibliographystyle{IEEEtran}	
	\bibliography{AnonymousSubmission/LaTeX/references}

\begin{thebibliography}{10}
\providecommand{\url}[1]{#1}
\csname url@samestyle\endcsname
\providecommand{\newblock}{\relax}
\providecommand{\bibinfo}[2]{#2}
\providecommand{\BIBentrySTDinterwordspacing}{\spaceskip=0pt\relax}
\providecommand{\BIBentryALTinterwordstretchfactor}{4}
\providecommand{\BIBentryALTinterwordspacing}{\spaceskip=\fontdimen2\font plus
\BIBentryALTinterwordstretchfactor\fontdimen3\font minus \fontdimen4\font\relax}
\providecommand{\BIBforeignlanguage}[2]{{%
\expandafter\ifx\csname l@#1\endcsname\relax
\typeout{** WARNING: IEEEtran.bst: No hyphenation pattern has been}%
\typeout{** loaded for the language `#1'. Using the pattern for}%
\typeout{** the default language instead.}%
\else
\language=\csname l@#1\endcsname
\fi
#2}}
\providecommand{\BIBdecl}{\relax}
\BIBdecl

\bibitem{iotddos2}
J.~Zhang, S.~Liang, F.~Ye, R.~Q. Hu, and Y.~Qian, ``Towards detection of zero-day botnet attack in iot networks using federated learning,'' in \emph{ICC 2023 - IEEE International Conference on Communications}, 2023, pp. 7--12.

\bibitem{iotddos1}
N.~Jaton, S.~Gyawali, and Y.~Qian, ``Distributed neural network-based ddos detection in vehicular communication systems,'' in \emph{2023 16th International Conference on Signal Processing and Communication System (ICSPCS)}, 2023, pp. 1--9.

\bibitem{llmabnormal}
\BIBentryALTinterwordspacing
J.~Zhu, S.~Cai, F.~Deng, B.~C. Ooi, and J.~Wu, ``Do llms understand visual anomalies? uncovering llm's capabilities in zero-shot anomaly detection,'' in \emph{Proceedings of the 32nd ACM International Conference on Multimedia}, ser. MM '24.\hskip 1em plus 0.5em minus 0.4em\relax New York, NY, USA: Association for Computing Machinery, 2024, p. 48–57. [Online]. Available: \url{https://doi.org/10.1145/3664647.3681190}
\BIBentrySTDinterwordspacing

\bibitem{ylu1}
L.~Li, J.~Li, C.~Chen, F.~Gui, H.~Yang, C.~Yu, Z.~Wang, J.~Cai, J.~A. Zhou, B.~Shen \emph{et~al.}, ``Political-llm: Large language models in political science,'' \emph{arXiv preprint arXiv:2412.06864}, 2024.

\bibitem{oct1}
W.~Chen, Z.~Li, and M.~Ma, ``Octopus: On-device language model for function calling of software apis,'' \emph{arXiv preprint arXiv:2404.01549}, 2024.

\bibitem{llmqw1}
J.~Xu, Z.~Li, W.~Chen, Q.~Wang, X.~Gao, Q.~Cai, and Z.~Ling, ``On-device language models: A comprehensive review,'' \emph{arXiv preprint arXiv:2409.00088}, 2024.

\bibitem{oct2}
W.~Chen, Z.~Li, S.~Xin, and Y.~Wang, ``Dolphin: Long context as a new modality for energy-efficient on-device language models,'' \emph{arXiv e-prints}, pp. arXiv--2408, 2024.

\bibitem{qunodllm}
J.~Xu, Q.~Wang, Y.~Cao, B.~Zeng, and S.~Liu, ``A general purpose device for interaction with llms,'' in \emph{Proceedings of the Future Technologies Conference}.\hskip 1em plus 0.5em minus 0.4em\relax Springer, 2024, pp. 613--626.

\bibitem{yl3}
L.~Gao, J.~Sherwood, N.~Aleisa, A.~Damoah, Y.~Lu, and X.~Qu, ``Human-centered ai agents for healthcare and education: A systematic literature review.''

\bibitem{qunddos}
S.~Verma, Q.~Wang, and E.~Bethel, ``Intelligent iot attack detection design via odllm with feature ranking-based knowledge base,'' \emph{arXiv preprint arXiv:2503.21674}, 2025.

\bibitem{weng2025thinking}
\BIBentryALTinterwordspacing
L.~Weng, ``Why we think,'' \emph{lilianweng.github.io}, 2025. [Online]. Available: \url{https://lilianweng.github.io/posts/2025-05-01-thinking/}
\BIBentrySTDinterwordspacing

\bibitem{wei2022chain}
J.~Wei, X.~Wang, D.~Schuurmans \emph{et~al.}, ``Chain of thought prompting elicits reasoning in large language models,'' in \emph{Advances in Neural Information Processing Systems (NeurIPS)}, 2022.

\bibitem{lewis2020retrieval}
P.~Lewis, E.~Perez, A.~Piktus \emph{et~al.}, ``Retrieval-augmented generation for knowledge-intensive nlp tasks,'' in \emph{Advances in Neural Information Processing Systems (NeurIPS)}, 2020.

\bibitem{snell2024scaling}
C.~Snell, J.~Lee, K.~Xu, and A.~Kumar, ``Scaling llm test-time compute optimally can be more effective than scaling model parameters,'' \emph{arXiv preprint arXiv:2408.03314}, 2024.

\bibitem{kojima2022large}
T.~Kojima, S.~S. Gu, M.~Reid, Y.~Matsuo, and Y.~Iwasawa, ``Large language models are zero-shot reasoners,'' \emph{Advances in neural information processing systems}, vol.~35, pp. 22\,199--22\,213, 2022.

\bibitem{grinsztajn2022tree}
L.~Grinsztajn, E.~Oyallon, and G.~Varoquaux, ``Why do tree-based models still outperform deep learning on typical tabular data?'' \emph{Advances in neural information processing systems}, vol.~35, pp. 507--520, 2022.

\bibitem{ollama}
F.~Liu, Z.~Kang, and X.~Han, ``Optimizing rag techniques for automotive industry pdf chatbots: A case study with locally deployed ollama models,'' \emph{arXiv preprint arXiv:2408.05933}, 2024.

\bibitem{llama3}
A.~Dubey, A.~Jauhri, A.~Pandey, A.~Kadian, A.~Al-Dahle, A.~Letman, A.~Mathur, A.~Schelten, A.~Yang, A.~Fan \emph{et~al.}, ``The llama 3 herd of models,'' \emph{arXiv preprint arXiv:2407.21783}, 2024.

\bibitem{gemma3}
G.~Team, A.~Kamath, J.~Ferret, S.~Pathak, N.~Vieillard, R.~Merhej, S.~Perrin, T.~Matejovicova, A.~Ram{\'e}, M.~Rivi{\`e}re \emph{et~al.}, ``Gemma 3 technical report,'' \emph{arXiv preprint arXiv:2503.19786}, 2025.

\bibitem{ciciot2023}
E.~C.~P. Neto, S.~Dadkhah, R.~Ferreira, A.~Zohourian, R.~Lu, and A.~A. Ghorbani, ``Ciciot2023: A real-time dataset and benchmark for large-scale attacks in iot environment,'' \emph{Sensors}, vol.~23, no.~13, p. 5941, 2023.

\end{thebibliography}
		
	%\end{thebibliography}	

\end{document}